\documentclass[epj]{svjour}
\usepackage{graphics}
\usepackage{bm}
\usepackage{placeins}
\usepackage[]{graphicx}
\usepackage{amsfonts}
\usepackage{bbm}
\usepackage{graphicx,color,amsmath,amssymb}
\usepackage{dsfont}

\def\blfootnote{\xdef\@thefnmark{}\@footnotetext}
\usepackage{amsmath, amssymb, graphics, setspace}

\newcommand{\mathsym}[1]{{}}
\newcommand{\unicode}[1]{{}}
\def\eg{{\em e.g.}}
\def\ie{{\em i.e.}}

\def\Tpc{{T_{\rm pc}}}
\newcommand{\beq}{\begin{equation}}
	\newcommand{\eeq}{\end{equation}}
\newcommand{\bea}{\begin{eqnarray}}
	\newcommand{\eea}{\end{eqnarray}}

\newcommand{\mlog}{\mathbb{L}\text{og}}

\begin{document}

\title{Spectral and Transport Properties of Quark-Gluon Plasma in a Nonperturbative Approach}
\titlerunning{Spectral and Transport Properties of QGP in a Nonperturbative Approach}
\author{Shuai Y.F. Liu\inst{1} \and Ralf Rapp\inst{1}
}                     

%
\institute{Department of Physics and Astronomy and Cyclotron Institute, Texas A\&M University,	College Station, TX 77843-3366, USA}
\date{Received: \today / Revised version:}
%
\abstract{
Nonperturbative methods play an important role in quantum many-body systems, 
especially in situations with an interplay of continuum and bound states and/or 
large coupling strengths between the constituents. 
Employing the Luttinger-Ward functional (LWF) we have computed the equation of state (EoS) 
of the quark-gluon plasma (QGP) using fully dressed selfconsistent 1- and 2-body 
propagators. We first give an alternative derivation of our previously reported results 
for resumming the ladder diagram series of the LWF using a ``matrix log" technique which 
accounts for dynamically formed bound and resonant states. 
Two types of solutions were found in selfconsistent fits to lattice-QCD data for the EoS, 
heavy-quark free energy and quarkonium correlators: a strongly coupled 
scenario (SCS) with broad parton spectral functions and strong meson resonances
near the transition temperature vs.~a weakly coupled scenario (WCS) with well-defined
parton quasiparticles and weak meson resonances. Here, we discuss 
how these solutions can be distinguished by analyzing the pertinent transport properties. 
We focus on the specific shear viscosity, $(4\pi )\eta/s$, and the heavy-quark diffusion 
coefficient, $(2\pi T) {\cal D}_s$, including its mass dependence. 
At low temperatures, in the SCS, they turn out to be a factor of 2 within their conjectured 
quantum lower bound, while they are a factor of 2-5 larger in the WCS. At higher
temperatures, the transport parameters of the two scenarios approach each other. 
We propose the ratio $ (4\pi\eta/s)/(2\pi T {\cal D}_s )$ as a measure
to distinguish the perturbative and strong-coupling limits of 5/2 and 1, respectively.
} 
%

\maketitle
\section{Introduction}
The theoretical investigation of the quark-gluon plasma (QGP) remains a challenging task in 
nuclear research. Of particular interest are its spectral and transport properties, which can be 
related to observables in ultrarelativistic heavy-ion collisions (URHICs). These quantities are 
not easily extracted from first-principles lattice-QCD (lQCD) computations~\cite{Ratti:2018ksb} 
which are performed in euclidean space-time. Various other theoretical tools, including 
Dyson-Schwinger, functional-renormalization group and Polyakov-Nambu-Jona-Lasinio 
approaches~\cite{Qin:2013ufa,Fischer:2017kbq,Khan:2015puu,Herbst:2013ufa,Fukushima:2017csk,Bastian:2018mmc}, 
are being pursued to meet this challenge. In the present work, we employ a thermodynamic $T$-matrix 
approach, which was originally designed to describe heavy quarks and quarkonia in the 
QGP~\cite{Cabrera:2006wh,Riek:2010fk}. More recently, it has been extended to the light sector 
to connect the in-medium properties of heavy quarks to those of the QGP bulk medium~\cite{Liu:2017qah}. 
The $T$-matrix approach utilizes a ladder resummation where the interaction kernel 
is based on a potential approximation while keeping the full dynamical information in the
one- and two-body spectral functions. This enables to directly calculate in real time 
and account for long-range nonperturbative interactions, thus facilitating insights into 
QGP properties that are complementary to other approaches as mentioned above. 
Among its benefits is the capability to evaluate various quantities, encompassing transport, 
spectral and bulk properties, within a rigorous many-body framework on the same footing. In
this way one can investigate interrelations between the microscopic structure of the QGP 
and its macroscopic properties as inferred from URHIC 
phenomenology~\cite{Heinz:2013th,Bernhard:2016tnd,Niemi:2015qia,Rapp:2018qla,Cao:2018ews}.


Within the $T$-matrix approach we employ a QCD-inspired Hamiltonian, where, in the spirit of 
an effective field theory, the inputs in form of an in-medium potential and bare masses 
(``Wilson coefficients") are matched to QCD utilizing lattice ``data" for quarkonium correlation 
functions, the HQ free energy, as well as the QGP equation of state (EoS). 
This is carried out in a selfconsistent Luttinger-Ward-Baym (LWB) 
formalism~\cite{PhysRev.118.1417,Baym:1961zz,Baym:1962sx} for the EoS, which accounts for the 
full off-shell dynamics of the in-medium particle propagators and scattering amplitudes 
in a conserving approximation.
In particular, the bound-state contribution to the EoS, encoded in the 
Luttinger-Ward functional (LWF), $\Phi$, is incorporated through a full resummation of the 
the $t$-channel ladder diagrams utilizing a numerical matrix-logarithm method. While we have 
employed this method before in a 3-dimensional (3D) reduced version~\cite{Liu:2016nwj,Liu:2017qah}, 
the present paper contains an alternative derivation in a full 4D context, although the practical 
application will still be carried out in a 3D-reduced form. 
The main new application in the present paper is the calculation of transport coefficients, 
specifically the shear viscosity of the bulk medium and the spatial HQ diffusion constant 
including a study of its HQ mass dependence. These will provide new insights into how they 
relate to each other and to the underlying spectral properties of partons and their 2-body 
correlations. 

Our paper is organized as follows. In Sec.~\ref{sec_form}, we briefly discuss the main 
components of the many-body formalism, containing the 4D derivation for the resummation 
of the LWF. In Sec.~\ref{sec_input}, we recall the inputs corresponding to 2 limiting 
scenarios (strongly coupled scenario (SCS) and  weakly coupled scenario (WCS)) as 
identified in our previous work, and summarize the key features of the pertinent spectral 
properties in Sec.~\ref{sec_spec}. In Sec.~\ref{sec_trans}, we calculate the transport 
coefficients for both coupling scenarios and propose their ratio as a novel measure to 
assess the coupling strength of the medium. In Sec.~\ref{sec_concl}, we summarize our main 
results and conclude.
\section{Many-Body Formalism}
\label{sec_form}
In this section, we lay out the many-body formalism in a self-contained form. 
In particular, the nonperturbative evaluation of the EoS is derived in a 4D version that goes
beyond our previous work~\cite{Liu:2017qah}. Although the practical application in the present 
paper will utilize a 3D reduced form, the framework can in principle be deployed to 4D 
approaches such as Dyson-Schwinger.

We start from the grand potential as a 
functional of the fully dressed single-particle propagator, $G$, 
\begin{align}
	\Omega (G)=\mp\text{Tr}\{\ln (-G^{-1}) +(G_0^{-1}-G^{-1}) G\}\pm\Phi(G) \ ,
	\label{Omega}	
\end{align}
where the trace (Tr) indicates 3-momentum ($\textbf{p}$) integrations and summations 
over Matsubara frequencies ($\omega_n$) and internal degrees of  freedom (\eg, spin 
and flavor); $G_0$ denotes the bare propagator. The single-particle selfenergy can be 
expressed as a functional derivative of the LWF, 
\begin{align}
	\Sigma (G)=\beta\delta\Phi(G)/\delta G \ ,
	\label{Sigma-f}	
\end{align}
where $\Phi(G)$ is usually constructed by a skeleton expansion to finite loop 
order~\cite{Blaizot:2000fc} ($\beta=1/T$ is the inverse temperature). 
Here we consider a situation where the selfenergy, $\Sigma (G)$, is calculated 
from a $T$-matrix amplitude with kernel $V$, with formal solutions  
\begin{align}
	&{T}=V+VGG{T}=(1-VGG)^{-1}V
	\label{Tmat}
	\\
	&\Sigma(G)={T} G=(1-VGG)^{-1}V G \  .
	\label{Sigma}		
\end{align}
Integrating Eq.~(\ref{Sigma-f}), together with 
Eq.~(\ref{Sigma}), yields 
\begin{align}
	& \Phi(G)=\frac{1}{\beta}\int dG (1-VGG)^{-1}V G=-\frac{1}{2\beta}\ln(1-VGG)\nonumber\\
	&=\frac{1}{2\beta}\{VG+\frac{1}{2}VGGVG+\cdots+\frac{1}{\nu}VGG\cdots VG\}G   \ .  
	\label{phi-op}		
\end{align}
The second line recovers the standard skeleton expansion, 
\begin{equation}
	\Phi(G)=\frac{1}{2}\text{Tr}\sum_{\nu=1}^\infty \frac{1}{\nu}\Sigma_\nu(G)G  \ ,
	\label{phi-skel}
\end{equation}
where $\Sigma_\nu(G)$ represents the $\nu^{\text{th}}$ order in the interaction in the skeleton diagram, 
expressed as a functional of $G$~\cite{PhysRev.118.1417}. 
The formally integrated form in the first line of 
Eq.~(\ref{phi-op}) is a nonperturbative expression which 
remains finite for large $V$, for which the perturbative series 
in the second line is usually divergent. This calls for a practical method to evaluate
the functional integral in Eq.~(\ref{phi-op}). 

Toward this end we resolve this expression into discrete 
energy-momentum indices (similar to the discretization of the $T
$-matrix integral equation~\cite{Haftel:1970zz}), and convert 
the log-function into a matrix representation. 
Considering 4-momentum as a single discretized variable, one can write the 2-body 
interaction kernel and single-particle propagators as 
\begin{align}
	&\mathbb{V}_{ij}\equiv V(\tilde{k}_i,\tilde{k}_j)
	\nonumber\\
	&\mathbb{G}_{ij}\equiv G(\tilde{k}_i)\delta_{ij} \ , \ 
	\mathbb{G}(\tilde{P})_{ij}\equiv G(\tilde{P}-\tilde{k}_i)\delta_{ij} \ ,
	\label{eq_VGdiscrete}		
\end{align}
where $\tilde{k}_{i,j}$=$(i\omega_n,\textbf{k})_{i,j}$ are relative in- and outgoing 
4-mo\-menta in Matsubara representation, and $\tilde{P}$ is the total 2-particle 4-momentum 
in the heat bath. With this notation one recovers the standard inverse-matrix 
solution~\cite{Haftel:1970zz} to the $T$-matrix as
\begin{equation}
	\mathbb{T}(\tilde{P})=[(\mathds{1}- \mathbb{VGG}(\tilde{P})]^{-1}\mathbb{V} \ , 
	\label{Tmat-inv}
\end{equation} 
with $T(\tilde{p},\tilde{q}|\tilde{p}',\tilde{q}')$ written as 
${T}(\tilde{P}-\tilde{k}_i,\tilde{k}_i|\tilde{P}-\tilde{k}_j,\tilde{k}_j)=\mathbb{T}(\tilde{P})_{ij}$.
We can now evaluate the logarithm in Eq.~(\ref{phi-op}) using a 
matrix representation, denoted by ``$\mlog$", 
as\footnote{A similar expression is known for the ground-state energy at zero 
	temperature~\cite{blaizot1986quantum} and for cold-atom systems ~\cite{PhysRevA.75.023610}.}

\begin{align}
	&\Phi=-\frac{1}{2}\int d^4\tilde{P}\ \text{Tr}\left\{\mathbb{L}\text{og}
	\left[\mathds{1}-\mathbb{VGG}(\tilde{P})\right]\right\}
	\ . 
	\label{phi-log}		
\end{align}
The ``Tr" of the matrix includes relative 4-momentum and internal degrees of freedom, 
and is followed by a scalar integration over $\tilde{P}$. Equation~(\ref{phi-log}) 
constitutes a practical formula to compute the ladder series of the LWF exactly; it is 
an example of a primitive functional of $\Sigma(G)$, \ie, a category of functional 
integrals which can be carried out by a matrix function.

We perform the $\mlog$ operation by recasting the LWF as 
\begin{equation}
	\Phi=\frac{1}{2}\int d^4\tilde{p} \ \text{ln}\Sigma(\tilde{p}) \ G(\tilde{p})  \ . 
	\label{phi-scalar}
\end{equation}
This can be seen by augmenting the $\mlog$ in Eq.~(\ref{phi-log}) 
with 1=$[\mathbb{GG}(\tilde{P})]^{-1} [\mathbb{GG}(\tilde{P})]$ and combining it 
with the first (inverse) factor to obtain
\begin{equation}
	\mlog \mathbb{T}(\tilde{P})=-{\mlog}\left[\mathds{1}- \mathbb{VGG}(\tilde{P})\right][\mathbb{GG}(\tilde{P})]^{-1} \ .
	\label{eq_lnTbykernel}         
\end{equation}
With $\mlog\mathbb{T}(\tilde{P})_{ij}=
\ln{T}(\tilde{P}-\tilde{k}_i,\tilde{k}_i|\tilde{P}-\tilde{k}_j, \tilde{k}_j)$,
we contract the diagonal forward-scattering $T$-matrix with 
$G$,  
\begin{equation} 
	\text{ln}\Sigma(\tilde{p})\equiv\int d^4\tilde{q} \ \text{ln}T(\tilde{p},\tilde{q}|\tilde{p},\tilde{q})\ G(\tilde{q})
	\label{eq_lnSelfEbylnT} \ , 
\end{equation}
which, together with the remaining propagator $G$, recovers Eq.~(\ref{phi-scalar}); note the 
formal similarity of Eq.~(\ref{phi-scalar}) with the skeleton expansion, Eq.~(\ref{phi-skel}). 
With this setup, the only change in going from the selfconsistent selfenergy, 
\begin{align}
	\Sigma (\tilde{p})=\int d^4\tilde{q} \ {T}(\tilde{p},\tilde{q}|\tilde{p},\tilde{q}) \ 
	G(\tilde{q}) \ ,
	\label{Sigma-int}              
\end{align}
to the LWF is replacing the inverse-matrix solution for $T$ from Eq.~(\ref{Tmat-inv}) 
by the matrix-logarithm, $\ln T$, from Eq.~(\ref{eq_lnTbykernel}) (and the factor 
$\mathbb{V}$ by $\mathbb{GG}^{-1}$). Standard techniques used to calculate the 
$T$-matrix, such as 3D reductions of the Bethe-Salpeter equation~\cite{Woloshyn:1974wm}, 
partial-wave expansions and center-of-mass (CM) 
approximations~\cite{Mannarelli:2005pz,Riek:2010py}, can be also be applied to 
calculate $\ln T$, cf.~also Refs.~\cite{Liu:2016nwj,Liu:2017qah}.

We deploy this formalism to the QGP within a Hamiltonian approach, systematically 
benchmarked by constraints from 
lQCD~\cite{Borsanyi:2010cj,Bazavov:2014pvz,Mocsy:2013syh,kaczmarek2007screening,Petreczky:2004pz}. 
Our main approximation is that the relevant interactions can be encoded in a 
potential-like driving kernel. 
Importantly, this allows to include remnants of the confining force above 
$T_{\rm pc}$, for which there is ample evidence from lQCD~\cite{kaczmarek2007screening,Bazavov:2014kva}. 
The effective Hamiltonian is of the form
\begin{align}
	H=&\sum \varepsilon(p)\psi^\dagger(\textbf{p})\psi (\textbf{p})+
	\hfill
	\nonumber\\
	&\frac{1}{2}\psi^\dagger(\frac{\textbf{P}}{2}-\textbf{p})\psi^\dagger(\frac{\textbf{P}}{2}+\textbf{p})V\psi(\frac{\textbf{P}}{2}+\textbf{p}')\psi(\frac{\textbf{P}}{2}-\textbf{p}') \ ,
	\label{eq_G1bare}		
\end{align}
where the summation is over momentum, spin, color and 
flavor ($N_f$=3 anti-/quarks plus gluons; for simplicity we assume 
spin degeneracy), and $\varepsilon(p)$=$\sqrt{M^2+\textbf{p}^2}$
with bare parton masses, $M$. The effective gluon is treated with a single-pole propagator in 
our approach. Thus, some expressions will differ by a factor of 1/2 compared to those using a
fully relativistic gluon propagator. For the 2-body potential we make the ansatz
\begin{align}
	V(\textbf{p},\textbf{p}')=\mathcal{F}^\mathcal{C}V_\mathcal{C}(\textbf{q})B(p,p')
	+\mathcal{F}^\mathcal{S}V_\mathcal{S}(\textbf{q})/R(p,p')
	\label{eq_potential}		
\end{align}
where $\textbf{q}=\textbf{p}-\textbf{p}'$ is the 3-momentum 
transfer, and $B$, $R$ are relativistic correction factors~\cite{Riek:2010fk}.
For the color factors, $\mathcal{F}^\mathcal{C,S}$, of the 
different two-body channels we use Casimir scaling for both Coulomb and string 
potentials, albeit with absolute values for the latter to maintain a strictly 
positive string tension~\cite{Petreczky:2004pz}.

The Hamiltonian is temperature dependent through the input bare masses and potentials. This implies a
modification of the standard thermodynamic relations as discussed in the appendix of 
Ref.~\cite{Liu:2017qah}, since we not only need to consider the ``evolution" of the states with 
temperature, but also the ``evolution" of the operators. This is similar to the case of a 
time-dependent Hamiltonian, where the evolution of both operators and states figure. The fundamental 
Hamiltonian is time and temperature independent. The time or temperature dependent effective 
Hamiltonian represent a macroscopic average of microscopic processes, such as quark/gluon 
condensate physics (giving rise to effective masses) which is beyond the scope of our current 
work. To recover the original thermodynamics requires a more explicit treatment of the averaged
microphysics in the masses and potentials, leading to an extra integral equation as suggested 
in the appendix in Ref.~\cite{Liu:2017qah}. We defer these studies to future work.

The in-medium potential is constrained following Ref.~\cite{Liu:2015ypa} by 
calculating the static HQ free energy, $F_{Q\bar Q}(r,T)$, 
as well as quarkonium correlator ratios~\cite{Cabrera:2006wh,Riek:2010fk,Riek:2010py} 
within the $T$-matrix formalism. This includes imaginary parts in the 2-body 
potential and the single-quark propagators calculated selfconsistently 
from the heavy-light $T$-matrix. 
In this way, an ansatz for the input potential, taken to be of in-medium 
Cornell-type~\cite{Megias:2005ve,Megias:2007pq},
\begin{equation}
	V_\mathcal{C}+V_\mathcal{S}=-\frac{4}{3}\alpha_s 
	\frac{e^{-m_d r}}{r}-\frac{\sigma e^{-m_s r- (c_b m_s r)^2}}{m_s} \ ,
	\label{eq_pot}
\end{equation}
is implemented into the $T$-matrix formalism, and its parameters 
($m_{d,s}$, $\alpha_s$, $\sigma$, $c_b$)~\cite{Liu:2017qah} are adjusted 
to lQCD data for $F_{Q\bar Q}(r,T)$ \cite{Mocsy:2013syh} and pseudoscalar 
quarkonium correlators~\cite{Aarts:2007pk,Aarts:2011sm}. An 
extra term, $-(c_b m_s r)^2$, in the exponential of the string 
interaction is introduced to better capture the residual effects 
of string breaking in the QGP. The screening mass of the string 
term, $m_s= (c_s m_d^2 \sigma/\alpha_s)^{1/4}$, is obtained from 
a one-loop calculation for the Debye mass of $V_\mathcal{S}$ 
(cf.~also Ref.~\cite{Burnier:2015tda}). The infinite distance 
limit of the color-singlet potential, 
$-\frac{4}{3}\alpha_s m_d + \frac{\sigma}{m_s}$, equals twice 
the Fock term for an individual static quark.
\begin{figure}[!t]
	\begin{center}
		\includegraphics[width=0.98\columnwidth]{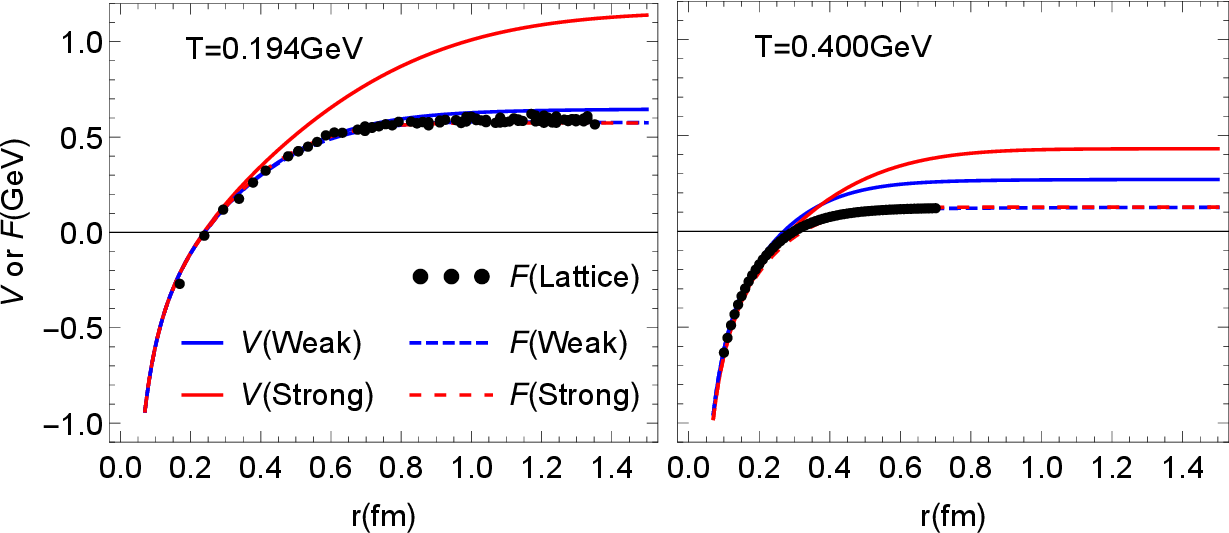}
		\caption{The SCS (red) and WCS (blue) input potentials (solid lines) and
			resulting free energies (dashed lines), compared to lQCD free energies at temperatures of 194\,MeV
			(left panel) and 400~MeV (right panel).}
		\label{fig_force}
	\end{center}
\end{figure}
\begin{figure*}[!t]
	\centering
	\includegraphics[width=2.0\columnwidth]{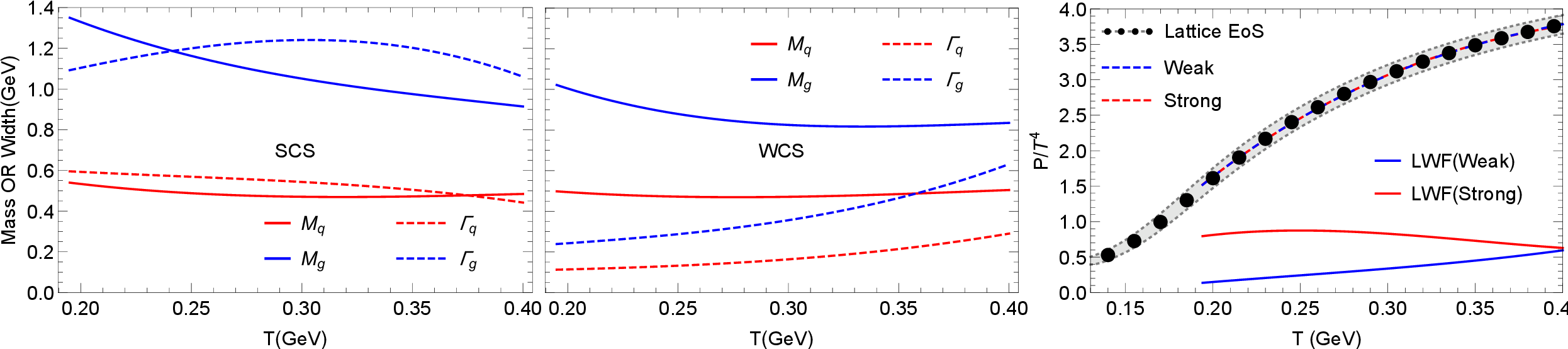}
	\caption{Temperature dependence of the fitted masses (solid lines) and emerging widths
		(dotted lines) for quarks (red lines) and gluons (blue lines) for the SCS (left panel) and
		WCS (middle panel), as figuring in the description of the pressure, $P/T^4$, from
		lQCD~\cite{Bazavov:2014pvz} (right panel).}
	\label{fig_EoS}
\end{figure*}
\section{Inputs for Two Scenarios}
\label{sec_input}
The solution that can simultaneously fit the three sets of lQCD data ($F_{\bar{Q}Q}$, 
quarkonium correlators, EoS) is not unique. In this section, we discuss two representative 
solutions in terms of the SCS, characterized by a rather strong potential (with long-range
remnants of the confining force in the QGP), and the WCS with a much weaker one close to the 
free energy itself, cf.~Fig.~\ref{fig_force}.
The parameter values for the SCS figuring in Eq.~(\ref{eq_pot}) from the selfconsistent fits are: 
$c_b$=1.3, $\alpha_s$=0.270, $\sigma$=0.225\,GeV$^2$, $m_d$=$-0.238$\,GeV + 2.915\,$T$, and 
$c_s$=0.01. Compared to previous work~\cite{Riek:2010fk}, the Casimir-scaled string term 
enables a better description of the color-octet free energy (not shown here). For the WCS we 
have: $c_b$=1.3, $\alpha_s$=0.270, $\sigma$=0.210\,GeV$^2$, 
$m_d$=$0.975~\text{GeV}- 0.135 \,\text{GeV}^2/T $, and $c_s$=0.1. 
These two quite simple parameterizations result in a rather good fit of the full 
numerical results to the lQCD data~\cite{Liu:2017qah}.


While the EoS is part of the overall nested fitting procedure, it involves more
directly the bare-quark ($M_q$) and -gluon ($M_g$) masses in the Hamiltonian, which are 
added to the selfconsistent Fock-masses. They provide the main handle to reproduce 
the lQCD data for the QGP pressure~\cite{Liu:2017qah}.
This is done by numerical iteration to selfconsistently solve the
$T$-matrices, Eq.~(\ref{Tmat-inv}), and selfenergies, Eq.~(\ref{Sigma-int}), 
for a trial mass value, and evaluate the matrix-log for the LWF $\Phi$ (including
all two-body color and flavor channels~\cite{Shuryak:2004tx} with spin degeneracy and 
angular momenta up to $l$=5, corresponding to 60 different $T$-matrices in the light 
sector~\cite{Liu:2017qah}.)
At low temperatures, $M_q/M_g$ approaches $C_F/C_A$=4/9, the ratio of Casimir 
factors for fundamental and adjoint representations in $SU_C$(3), reflecting 
the infinite-distance limits of the static potential. At high temperature, the 
ratio $M_q/M_g$ approaches the perturbative value, $\sqrt{1/3}/\sqrt{3/4}$~\cite{Levai:1997yx}. 
This setup for the masses thus mimics a dynamical mass generation from confinement (via
the 1-body selfenergy from the long-distance limit of the confining force), 
see Ref.~\cite{Liu:2017qah} for further details.

The resulting ``thermal" quark and gluon masses are comparable in magnitude to those 
extracted from quasiparticle model fits to the 
EoS~\cite{Peshier:2005pp,Levai:1997yx,Plumari:2011mk,Berrehrah:2016vzw}, 
although the temperature dependence found here is weaker, especially for the quark masses at 
temperatures approaching $\Tpc$ (from above), cf.~the left and middle 
panels of Fig.~\ref{fig_EoS}.
The gluon masses in the SCS are slightly larger than in the WCS, as they are selfconsistently 
related to the potentials as mentioned above~\cite{Liu:2017qah}.
While the parton masses are comparable between the SCS and WCS, the deviations between their 
potentials are much more significant, which will lead to quite different predictions 
for the spectral and transport properties, as will be discussed in the following two sections.
\begin{figure*}[t]
	\begin{center}
		\includegraphics[width=0.99\columnwidth]{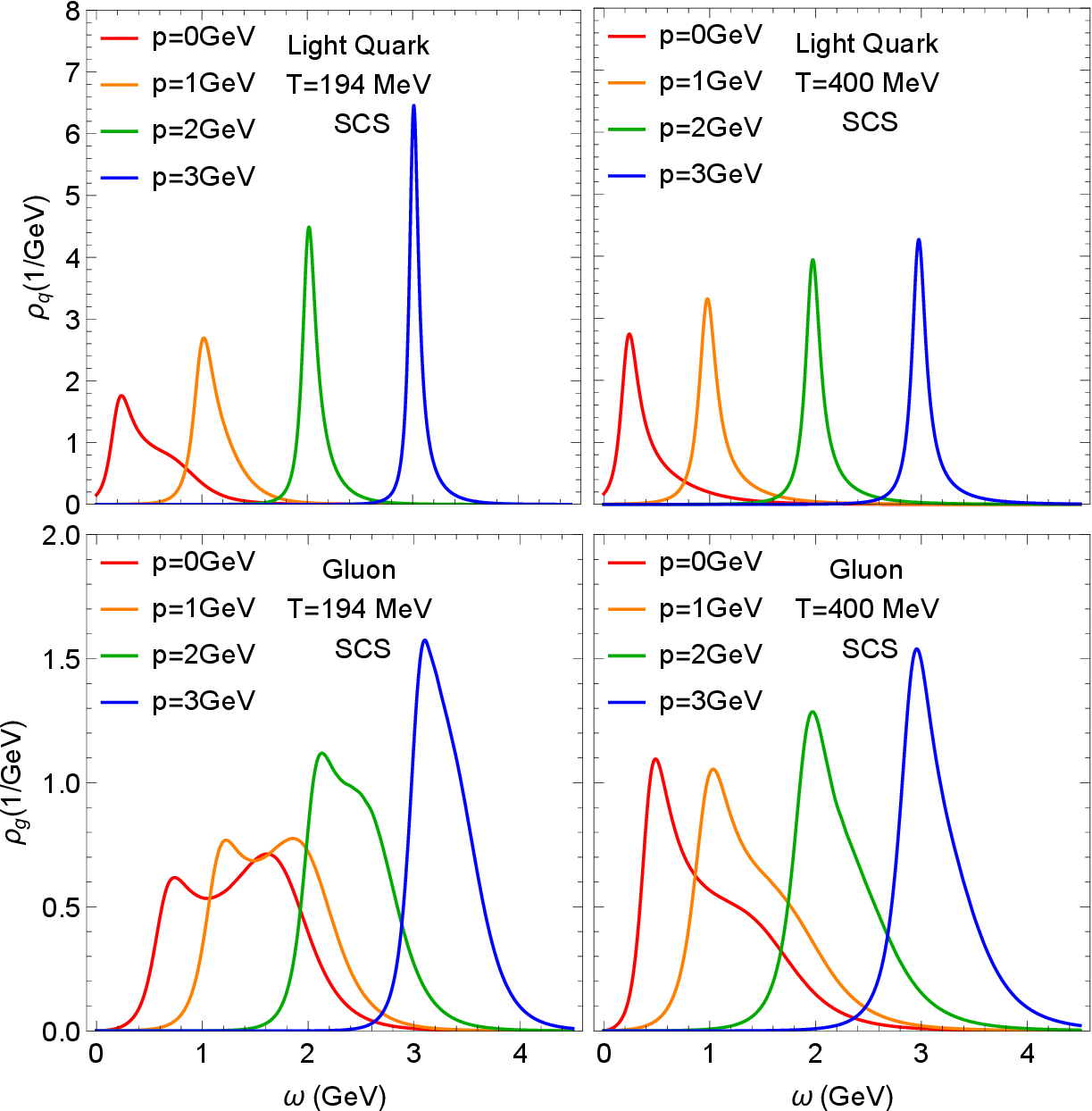}
		\includegraphics[width=0.99\columnwidth]{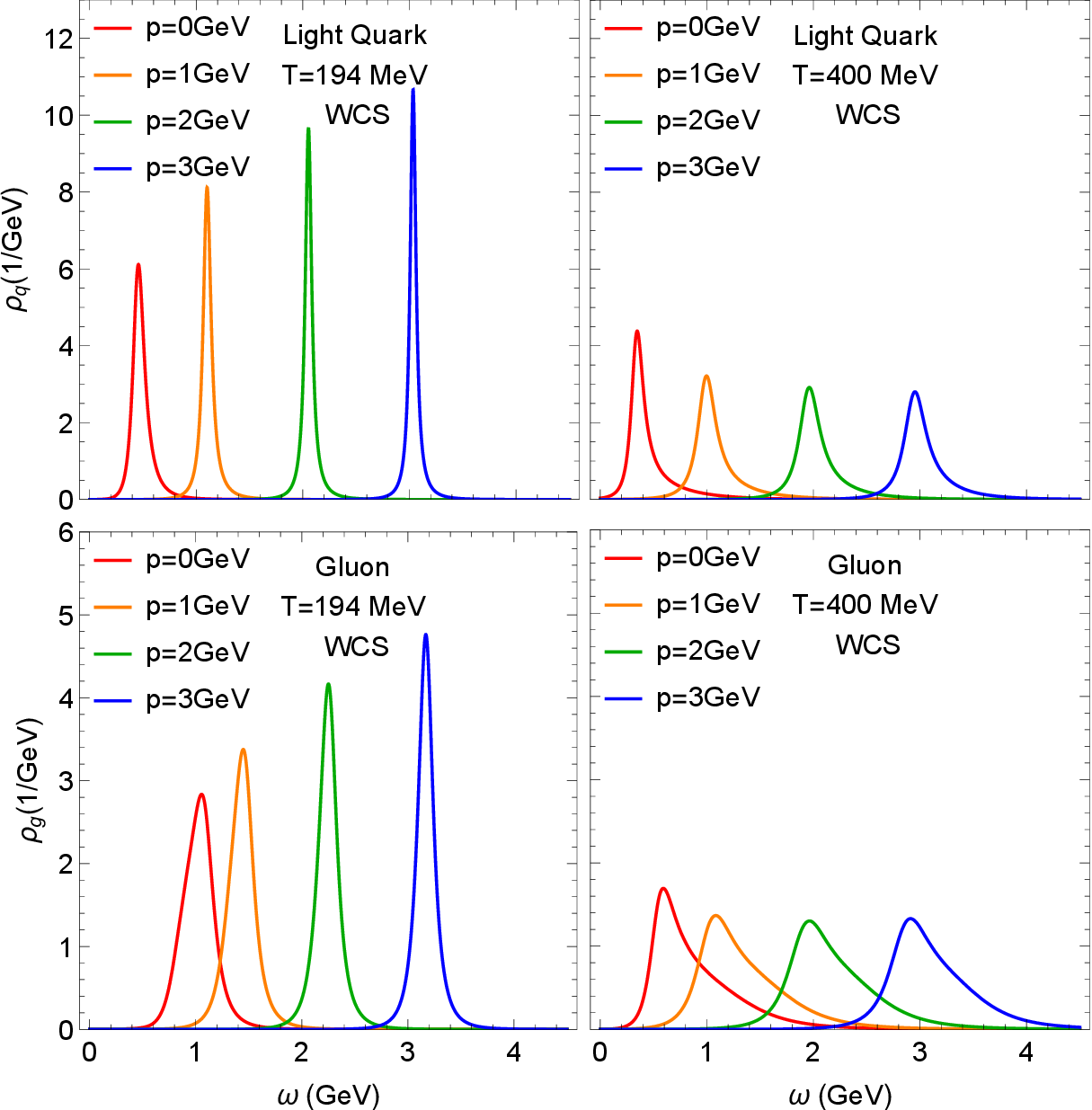}
		\caption{In-medium spectral functions for quarks (upper panels) and gluons (lower panels)
			in the SCS (left panels, for $T$=194\,MeV and T=400\,MeV) and WCS (right panels, for
			$T$=194\,MeV and T=400\,MeV) for 3-momenta $p$=0,1,2,3\,GeV (color coding).}
		\label{fig_Aparton}
	\end{center}
\end{figure*}

\begin{figure*}[!t]
	\begin{center}
		\includegraphics[width=2\columnwidth]{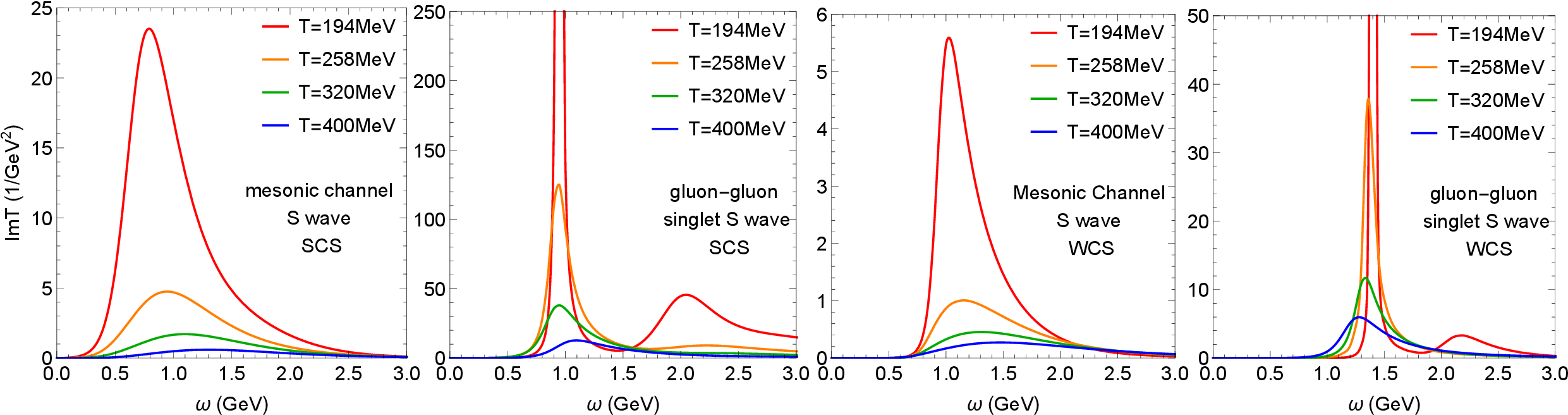}
		\caption{Two examples of imaginary parts of two-body $T$-matrices (of a total of 60 included
			in the LWF), for the color-singlet $S$-wave $q\bar q$ and $gg$ channels for the SCS (left panels)
			and WCS (right panels) at different temperatures (color-coded).}
		\label{fig_ImT}
	\end{center}
\end{figure*}

\section{Spectral Properties of Two Scenarios}
\label{sec_spec}
In this section, we focus on the emerging spectral properties within the two solutions, 
first for the SCS in Sec.~\ref{ssec_scs} and then for the WCS in Sec.~\ref{ssec_wcs}.

\subsection{SCS}
\label{ssec_scs}
The increasing role of the LWF, which encodes the bound-state contributions~\cite{Rapp:1993ih}, 
as $\Tpc$ is approached from above suggests a transition from parton quasiparticles to hadronic 
degrees of freedom (see right panel of Fig.~\ref{fig_EoS}). This is a dynamical transition 
driven by an increasing interaction strength (due to a reduced screening in the 
potential) which is very different from quasiparticle models where parton degrees of freedom 
are suppressed kinematically by an increasing mass. Our interpretation is 
corroborated upon inspecting the selfconsistent spectral functions implicit in the 
calculation, plotted in Fig.~\ref{fig_Aparton} for partons and in Fig.~\ref{fig_ImT}
for their two-body interaction (imaginary part of $T$ matrices). At low temperatures and 
3-momenta, the light-quark spectral functions are strongly broadened with their strength 
spread over an energy range of about 1\,GeV (cf.~left panels in Fig.~\ref{fig_Aparton}). 
This is well above their mass values, implying a loss of quark quasiparticles at small 
wavelengths in the QGP, again very different from quasiparticle-type 
models~\cite{Levai:1997yx,Plumari:2011mk,Berrehrah:2016vzw} where the quark widths are 
zero or small compared to their masses. 
In addition, large negative real parts of the low-energy 
selfenergy, $\text{Re}\Sigma(\omega,p)$, generate a low-energy solution for 
the dispersion relation, $\omega-\sqrt{M^2+p^2}-\text{Re}\Sigma(\omega,p)\sim0$, 
corresponding to a collective mode caused by off-shell interactions through near-threshold 
resonances in the $T$-matrix bound states which emerge at low temperature, cf.~left panels 
in Fig.~\ref{fig_ImT}. 
The situation is similar for gluons (lower left panels of Fig.~\ref{fig_Aparton}) and their 
interaction amplitudes (second panel in Fig.~\ref{fig_ImT}). Their spectral distributions 
also develop low-energy collective modes 
well below the nominal thermal gluon mass and 2-gluon threshold, respectively. 
However, the large mass of gluons near $\Tpc$ suggests that they largely decouple from the 
hadronization process of the system.

Despite the large increase in the degrees of freedom with temperature, the parton widths 
$\Gamma_{q,g}$ vary little in the SCS, cf.~left panel in Fig.~\ref{fig_EoS}. This is quite 
different from the perturbative expectation, $\Gamma \propto g^2T$, and another manifestation 
of the increase in interaction strength as $T\to \Tpc$ from above. This increase 
is largely driven by the formation of pre-hadronic resonances as discussed above. 
With increasing temperature, the resonances dissolve, which, in turn, leads to better 
defined partonic quasiparticles. In this sense, our selfconsistent approach exhibits 
a smooth transition from hadronic to quark degrees of freedom as temperature (or 
3-momentum) increases. 
We also note that the color-singlet $q\bar q$ $S$-wave bound-state mass of $\sim$0.8\,GeV 
near $\Tpc$ (left panel of Fig.~\ref{fig_ImT}) is intriguingly close to the vacuum 
$\rho$-meson mass, with a broad spectral function not unlike results from in-medium hadronic 
calculations~\cite{Rapp:2009yu}. 
Its large width suggests that the dissociation and regeneration rate of hadronic states 
is quite large, which may give a microscopic justification of why statistical hadronization 
is a good approximation for the dynamical hadronization within a transport approach.

\subsection{WCS}
\label{ssec_wcs}
In the WCS, the interaction contribution to the total pressure represented by the LWF remains 
quite low even as the temperature approaches $\Tpc$, cf.~right panel of Fig.~\ref{fig_EoS}.
This suggests that there is no transition in the degrees of freedom of the system, which
is supported by the spectral functions of the light partons remaining quasiparticle-like 
down to low temperatures (cf.~right panels in Fig.~\ref{fig_Aparton}). Indeed, the pertinent 
widths, $\Gamma_{q,g}$, shown in middle panel of Fig.~\ref{fig_EoS}, stay well below their 
masses. Their increase with temperature 
is slightly faster than linear, presumably caused by the slower-than-linear 
increase of the screening masses figuring in the WCS potential.

The strength of two-body $T$-matrix amplitudes is much reduced in the WCS, by roughly a factor of
5 at the lowest temperature, cf.~the two right panels in Fig.~\ref{fig_ImT}. The small amplitudes lead 
to small collision rates of the partons which is the reason for maintining quasiparticle-like spectral 
functions at all temperatures and 3-momenta.
The resonance peak of the $S$-wave mesonic channel is at around 1~GeV, just above the two-particle 
threshold. The latter feature is probably the main reason why the width of the threshold resonances 
at small temperature is not dramatically smaller than in the SCS. 

\section{Transport Properties of Two Scenarios}
\label{sec_trans}

\begin{figure}[t]
	\begin{center}
		\includegraphics[width=0.99\columnwidth]{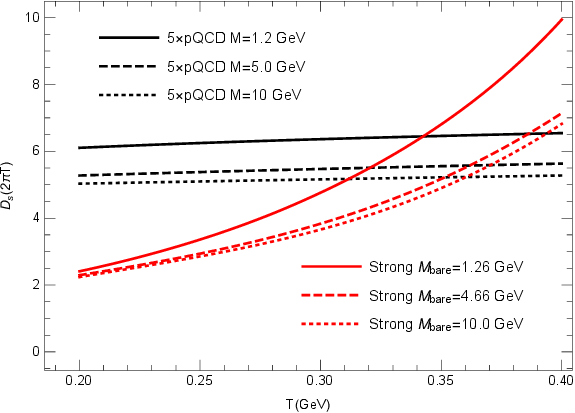}
	\end{center}
	\caption{Temperature dependence of the dimensionless HQ diffusion coefficient, ${\cal D}_s(2\pi T)$,
		for pQCD Born calculations (including an overall factor $K$-factor of 5 in the relaxation rate; 
		black lines) and the SCS (red lines). The solid, dashed and dotted lines are for $c$, $b$ and $b^+$
		quarks, respectively. The in-medium mass $\Delta M (T)$ of SCS is discussed in Ref~\cite{Liu:2017qah}
	}
	\label{fig_ds}
\end{figure}

\begin{figure*}[t]
	\begin{center}
		\includegraphics[width=0.99\columnwidth]{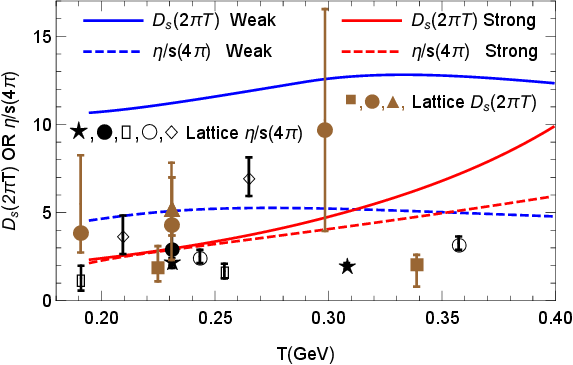}
		\includegraphics[width=0.99\columnwidth]{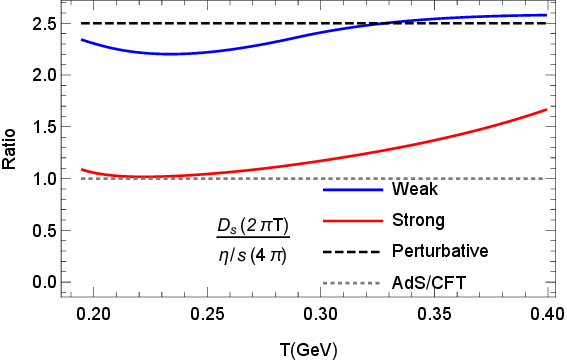}
	\end{center}
	\caption{Left panel: Specific viscosity, $4\pi\eta/s$ (dashed lines), and charm-quark diffusion
		coefficient scaled by the thermal wavelength, ${\cal D}_s(2\pi T)$ (solid lines) for the SCS
		(red lines) and WCS (blue lines), compared to lQCD data
		(\{$\bigstar$~\cite{Pasztor:2018yae}, {\large$\bullet$}~\cite{Mages:2015rea},
		$\square$~\cite{Meyer:2007ic}, $\bigcirc$~\cite{Meyer:2009jp}, $\Diamond$~\cite{Nakamura:2004sy}\}
		for $4\pi\eta/s$ in black and \{$\blacksquare$~\cite{Ding:2012sp},
		{\large$\bullet$}~\cite{Banerjee:2011ra},  $\blacktriangle$~\cite{Francis:2015daa}\}
		for ${\cal D}_s(2\pi T)$ in brown).
		Right panel:the ratio $[(2\pi T){\cal D}_s]/[4\pi\eta/s]$ for SCS (red solid line) and WCS
		(blue solid line) compared to perturbative (dashed line) and AdS/CFT (dotted line) limits.}
	\label{fig_coeff}
\end{figure*} 

We now utilize our results to compute two prominent transport coefficients of the QGP, the 
spatial heavy-quark diffusion coefficient, ${\cal D}_s$, and the shear viscosity $\eta$. 
Since temperature is a macroscopic external parameter for time evolution in the linear 
response theory, and since the latter does not affect the temperature or other macroscopic 
properties of the background medium, the usual theoretical tools for evaluating transport 
coefficients, such as the Kubo formula, should still work for our temperature-dependent 
Hamiltonian.

For ${\cal D}_s$, we employ the formalism of Refs.~\cite{Riek:2010fk,Riek:2010py} with an 
additional off-shell extension~\cite{Danielewicz:1982kk} to account for the quantum 
effects of the HQ spectral functions as discussed in Ref.~\cite{Liu:2018syc}. 
Schematically, the pertinent friction coefficient can be expressed as
\begin{align}
	A(p)=&\left\langle (1-\frac{\textbf{p}\cdot\textbf{p}'}{p^2})\rho_i\rho_i\rho_Q\right\rangle
\end{align}
where $\textbf{p}(\textbf{p}')$ is the incoming (outgoing) HQ quark 
momentum and the $\rho_{i(Q)}$ are the light-parton (HQ)
spectral functions. The diffusion coefficient follows from the zero-momentum
limit of the friction coefficient, ${\cal D}_s= T/(A(p=0) m_Q)$, where $m_Q$ denotes
the HQ mass. In addition to its temperature dependence, the HQ mass dependence of ${\cal D}_s$
is of considerable interest. A mass {\em in}dependence would be an important feature 
to consider ${\cal D}_s$ as a universal coefficient of the medium. Indeed, the leading mass  
dependence is divided out in its relation to the HQ relaxation rate, $A(p=0)$, whose decrease
with $m_Q$ simply reflects the kinetic property that heavier quarks take longer to thermalize.
We have therefore conducted calculations for charm quarks, bottom quarks and a still heavier 
quark (which we denote by $b^+$) with about twice the $b$-quark mass. 
In Fig.~\ref{fig_ds} we plot our results for 
$2\pi T {\cal D}_s$, scaled dimensionless by the thermal wavelength of the medium $1/2\pi T$. 
Let us  first inspect the results of time-honored perturbative QCD (pQCD) calculations using the 
leading-order Born diagrams for HQ scattering off thermal partons~\cite{Svetitsky:1987gq} (also
including an overall $K$ factor which is of no relevance in the present discussion). They are 
essentially constant with temperature and vary only weakly with $m_Q$, by up to 15\%. The $m_Q$
dependence suggests a saturation for large $m_Q$, thus converging toward a universal behavior
(which is reasonably well realized already by $b$ quarks). 
In the SCS, we observe a rather pronounced temperature dependence reflecting the reduced interaction
strength in the QGP as $T$ increases. However, the $m_Q$ dependence also exhibits nontrivial behavior, 
with a near degeneracy at low $T$ which is lifted as $T$ increases. The spread in $m_Q$ remains
small for $b$ and $b^+$ quarks even at higher $T$, indicating a good universality in $m_Q$, 
while the latter is broken for $c$ quarks indicating that $m_c/T$ is becoming too small. The genuine
(universal) temperature dependence in the HQ interaction strength is thus represented by the $b$ 
and $b^+$ quark curves. This reiterates the importance of high-precision bottom transport 
observables in URHICs.     

For $\eta$, we employ the standard Kubo formula using the leading-density 
energy-momentum tensor~\cite{zubarev1974nonequilibrium} with relativistic 
extension, 
\begin{align}
	\eta=&\lim\limits_{\omega\rightarrow 0}\sum_i\frac{\pi d_i}{\omega} \int
	\frac{d^3\textbf{p}d\lambda}{(2\pi)^3} \frac{p_x^2p_y^2}{\varepsilon^2_i(p)}
	\rho_i(\omega+\lambda,p)\rho_i(\lambda,p)\nonumber\\&\times[n_i(\lambda)-n_i(\omega+\lambda)]
	\ , 
\end{align}
where the $n_i(\omega)$ are the thermal distribution functions of light partons and 
$d_i$ the pertinent degeneracies. Higher-order corrections are expected to be 
small~\cite{Iwasaki:2006dr,Iwasaki:2007iv,Lang:2013lla,Ghosh:2014yea,Christiansen:2014ypa}. 
For example, in Refs.~\cite{Ghosh:2014yea,Enss:2010qh} second-order contributions have been 
found with a parametric dependence of $1/\Gamma^2$ (with $1/\Gamma$ for the leading loop). 
This suggests that for strongly coupled systems, with large particle widths, higher-order 
contributions are suppressed. A more systematic study to include higher-order corrections 
to the viscosity as discussed in Ref.~\cite{Enss:2010qh} will be reported elsewhere.

Before discussing the numerical results of the SCS and WCS, let us recall two limiting 
cases~\cite{Rapp:2009my} which will be instructive for comparisons with our results. 
In the strongly coupled limit, AdS/CFT approaches have been used to calculate the 
transport coefficients of interest here. The spatial HQ diffusion coefficient was found 
at ${\cal D}_s \approx1/(2\pi T)$~\cite{Gubser:2006bz,CasalderreySolana:2006rq,Gubser:2006qh}. 
Also, the shear viscosity is 
conjectured to reach a universal strongly-coupled limit of 
$\eta/s=1/(4\pi)$~\cite{Kovtun:2004de,Buchel:2004di}, 
where $s$ denotes the entropy density. This suggests a relation
\begin{align}
	(2\pi T {\cal D}_s)\approx 1\times(4\pi\eta/s)
	\label{eq_ratios}
\end{align}
in a strongly-coupled system. 
For a weakly-coupled massless gas, the viscosity can be evaluated in a classical kinetic 
theory as~\cite{Danielewicz:1984ww}
\begin{align}
	\eta/s\approx (\frac{4}{15}n\langle p\rangle\lambda_{\text{tr}})/s =\frac{1}{5}T\lambda_{\text{tr}}
\end{align}
with $ n\langle p\rangle=\varepsilon $ and $ Ts =\varepsilon+P=4/3\,\varepsilon$. Here, $n$, 
$\varepsilon$ and $P$ denote the particle density, energy density and pressure, respectively. 
Using a momentum transfer mean-free-path of  
$ \lambda_{\text{tr}}\approx\tau_{\text{tr}}\approx\tau_Q m_Q/T={\cal D}_s $ where $ \tau_{\text{tr}}$ 
and $\tau_Q$ are mean-free-time and HQ relaxation time, we obtain a relation
\begin{align}
	(2\pi T {\cal D}_s)\approx 5/2\times(4\pi\eta/s)
	\label{eq_ratiow}
\end{align}
for a weakly-coupled system. 

In the left panel of Fig.~\ref{fig_coeff} 
we plot the dimensionless-scaled transport coefficients together with the pertinent lQCD 
data.  The hierarchy between SCS and WCS basically reflects the interaction strength of 
the two scenarios, with the former leading to smaller values due to larger scattering rates. 
For the SCS at low temperatures, both coefficients are at approximately twice the value for 
the conjectured lower quantum bounds. It is interesting to note that the relatively close 
proximity to the quantum bounds is accompanied by the loss of (long-wavelength) quasiparticles 
in the system~\cite{Kovtun:2004de,Enss:2010qh}, as discussed in the previous section. 
For the WCS, the low-temperature $c$-quark diffusion coefficient is about a factor of 5 larger than 
in the SCS, but only about a factor $\sim$2 for the viscosity. Overall,the lQCD data favor the 
SCS.  In the SCS, there is a clear trend for the coefficients to increase with temperature, 
toward a more weakly-coupled medium. However, from inspecting the individual coefficients 
alone, it is not easy to tell what really constitutes the notion of a weakly- or 
strongly-coupled medium. 

We therefore propose the ratio $[(2\pi T){\cal D}_s]/[4\pi\eta/s]$ as a quantity to better distinguish  
strongly- and weakly-coupled media, cf.~the right panel of Fig.~\ref{fig_coeff}. As argued with 
Eq.~(\ref{eq_ratiow}) and Eq.~(\ref{eq_ratios}), this ratio is expected to be 1 in the strongly- 
and 5/2 in the weakly-coupled limit. Interestingly, this ratio is indeed close to 
one in the SCS at low temperature, increasing toward higher temperatures, yet still
significantly below 5/2 even at $T$=400\,MeV (where hydrodynamics is still believed to work well in
URHICs); we recall, however, that (part of) this increase is presumably due to the relatively small
$c$-quark mass; it is less in the HQ limit. On the other hand, the ratio in the WCS is close to 5/2 
even at low temperature (despite the rather small specific viscosity), with insignificant temperature 
dependence.

\section{Conclusions}
\label{sec_concl}
In the present work we have utilized our earlier developed nonperturbative quantum 
many-body approach to the QGP to analyze its transport properties. First, we have 
rederived in full 4D the selfconsistent all-order resummed Luttinger-Ward functional, 
which plays a key role in the description of QGP structure especially at low 
temperatures, where it accounts for dynamically formed bound states as a mechanism
for a gradual transition from partonic to hadronic degrees of freedom.    
After briefly reviewing the essential features of two solutions in fits to lQCD data 
for HQ and bulk properties (EoS), corresponding to a ``strong" and ``weak" underlying 
potential scenario (SCS and WCS), we have calculated pertinent transport coefficients, 
in particular the specific shear viscosity and the HQ diffusion coefficient. 
For the latter, we have investigated its dependence on the HQ mass and found that for
bottom quarks it is close to a universal limit for large $m_Q$, while for charm quarks it
can be significantly larger, especially toward higher temperatures.
In the SCS, it turns out that both the specific shear viscosity and the scaled HQ diffusion 
coefficient are at about 2 times their conjectured lower quantum bound at 
temperatures near $\Tpc$, and gradually increasing with temperature. The results of the SCS 
are in a favorable range for URHIC phenomenology using hydrodynamic bulk evolution models 
for light hadron spectra and HF transport approaches for HF observables. On the other hand, 
in the WCS, the transport parameters are markedly larger, in a range which is incompatible 
with experimental observations. Thus, the transport coefficients, when put into context with 
URHIC phenomenology, can discriminate the SCS and WCS solutions, which was not possible on 
the basis of the fits to lQCD data alone. We have then proposed the ratio of the HQ 
diffusion coefficient over specific viscosity as a more quantitative measure of the 
notion of a weakly vs.~a strongly coupled medium, with limiting values of 5/2 vs.~1, 
respectively. Somewhat surprisingly, the results for the WCS turn out to be close to 
the former over the considered temperature range, while the SCS results are close to 
one near $\Tpc$. Recalling the underlying spectral properties of the two scenarios, 
with the melting of parton quasiparticle structures in the SCS at large wavelength 
and low $T$, while quasiparticles prevail in the WCS at all $p$ and $T$, we corroborate 
earlier qualitative expectations along these lines. We also recall that the key agent 
in the strong coupling properties are long-range remnants of the confining force (which 
are prominent in the SCS but not in the WCS potential), thus suggesting an intimate 
connection between confinenment and the strongly coupled QGP.

\section*{Acknowledgements} 
This work has been supported by the U.S. NSF through grant nos. PHY-1614484 and PHY-1913286.

\bibliographystyle{h-elsevier}
\bibliography{refcnew}

\end{document}